\documentclass[%
 preprint,
 amsmath,amssymb,
 aps,
]{revtex4-1}

\usepackage{graphicx}
\usepackage{dcolumn}
\usepackage{subcaption}
\usepackage{hyperref}
\usepackage{bm}

\newcommand{\bs}[1]{{{{\bf #1}}}}
\newcommand{\bos}[1]{\boldsymbol{#1}}

\newcommand{\Eq}[1]{Eq.\,(\ref{#1})}

\newcommand{\Ref}[1]{Ref.~\onlinecite{#1}}
\newcommand{\Refs}[1]{Refs.~\onlinecite{#1}}
\newcommand{\p}{^{\prime}}
\newcommand{\iu}{{i\mkern1mu}}
\newcommand{\dx}{{\scriptstyle \Delta} x}

\newcommand{\mdx}{m\,{\scriptstyle \Delta} x}

\newcommand{\bra}[1]{\langle #1\vert}
\newcommand{\ket}[1]{\vert#1\rangle}
\newcommand{\braket}[2]{\langle#1|#2\rangle}

\newcommand{\diag}{\textrm{diag}}

\newcommand{\lrb}[1]{\left( #1\right)}

\begin{document}

\preprint{APS/123-QED}

\title{
Solver for the electronic V-representation problem of time-dependent density functional theory}

\author{James Brown}
\author{Jun Yang}%
\author{James D Whitfield}%
 \email{James.D.Whitfield@Dartmouth.edu}
\affiliation{%
 Dartmouth College\\
}%

\date{\today}

\begin{abstract}
    One route to numerically propagating quantum systems is time-dependent density functional theory (TDDFT). The application of TDDFT to a particular system's time evolution is predicated on $V$-representability which we have analyzed in a previous publication. 
    Here we describe 
    a newly developed solver for the scalar time-dependent Kohn-Sham potential. 
We present and interpret the force-balance equation central to our numerical method, describe details of its implementation, and present illustrative numerical results for one- and two-electron systems. A new characterization of $V$-representability for one-electron systems is also included along with possible improvements and future directions.
\end{abstract}

\maketitle

\tableofcontents
\begin{enumerate}
    \item $\dx$, Spacing between basis functions in position space.
    \item $m$, Index for location of phase-space basis function in position space at $m\dx$, not necessarily an integer but $m$ has integer spacing.
    \item  $ \vec{m}=\left[m_1,m_2,...m_D\right] $, The vector of all one-dimensional indices for the multidimensional basis function with $D$ dimensions.
    \item $L$, Total number of one dimensional phase space functions.
    \item $N$, Size of basis.
    \item $\hat{H}$, operators have the hat symbol.
    \item $\bs{H}$, Matrices will be capital letters in bold and non-italicized. 
    \item $\bs{v}$, Vectors will be small letters in bold and non-italicized.
    \item $\bs{v}^{\dagger}$ is the conjugate transpose of $\bs{v}$
    \item $V^{(i)}$: the $(i)$ superscript refers to a specific particle.
\end{enumerate}

\section{\label{sec:intro}Introduction}
Important classes of time-evolution algorithms widely employed by chemists and physicists are based on reduced descriptions of the wave function. These include methods focused on the two-body reduced density matrix and the electron density (the diagonal of the one-body reduced density matrix). 

Since all interactions of non-relativistic Hamiltonians are between at most two electrons, the $N$-electron wave function, $\Psi$, contains more information than necessary.  For this reason, the two-electron reduced density matrix (2RDM) contains enough information to characterize properties of non-relativistic quantum systems~\cite{Mazziotti11}. However to minimize the energy, one must characterize the set of 2RDMs corresponding to a valid $N$-electron wave function otherwise nonphysical 2RDMs allow unbounded solutions.  The characterization of the 2RDMs that came from $N$-electron wave functions is known as the \textit{$N$-representability problem}~\cite{Coulson60}.  The $N$-representability problem was proven to be QMA-complete~\cite{Lui07} highlighting the theoretical difficulty of 2RDM methods.  Nonetheless, there has been successful efforts to perform time evolution using 2RDM methods~\cite{Lackner15}.

An even more concise description is afforded by the ground state one-electron probability density, 
\begin{equation}
	n_t=\textrm{diag}\left(\textrm{Tr}_{2\ldots N}\ket{\Psi_t}\bra{\Psi_t}\right)
\label{eq:density}
\end{equation}
which we will simply refer to as the density.  The Hohenberg-Kohn theorems dictate that the ground state density, $n_{gs}$, is sufficient to characterize all properties of the quantum system~\cite{Hohenberg64}.  This provides the basis for density functional theory (DFT).  While theoretically compelling, many functionals to efficiently compute properties from the density are unknown. Moreover, the universal functional necessary for evaluating the 
energy is unlikely to be determined numerically even to only polynomially accurate in the size of the system. Despite the numerous approximations to the universal functional, computational complexity arguments \cite{Schuch09,Whitfield2014} showed that obtaining the numerically exact functional is intractable even with quantum computation.  It may be possible to obtain a functional using a particular form of the Kohn-Sham orbitals.\cite{Hatcher2019} The corresponding time-dependent result~\cite{Whitfield14a} states, for sufficiently well-behaved systems, the potential can be computed efficiently with access to a quantum computer. Unlike the ground state result, the time-dependent complexity analysis relied on the Kohn-Sham (KS) construction lying at the heart of nearly all practical schemes for DFT.  In this letter, we return to the analysis begun in our previous work~\cite{Whitfield14a} using a combination of theory and numerics.

The Kohn-Sham (KS) construction considers a system of non-interacting electrons that can be used to replace a given interacting system of electrons. An electronic system with with two-body interactions, $\hat W$ is governed by Hamiltonian $\hat {\bf T}+\hat {\bf V}+\hat {\bf W}$.  The KS system is a new collection of non-interacting electrons with a new additional potential term $\hat {\bf V}^{KS}$. The new potential in the KS Hamiltonian, $\hat {\bf T}+\hat {\bf V}+\hat {\bf V}^{KS}$, is selected such that the one-electron probability density of the non-interacting system matches that of the given interacting system of interest.

Let us define the general $V$-representation problem of a KS system as the task of constructing a model system which has the same expectation values on selected observables as a target system. $V$-representability refers to the existence of solutions to this problem when different constraints are placed on the model system.  The time-dependent subset of $V$-representation problems considers as input an initial state and the target trajectory of selected observables, and the task is to find the correct time-dependent fields for a specified control Hamiltonian. This general framework is not limited to electronic systems as illustrated by a study of this problem in the context of spin systems  \cite{Tempel2012}.  Here, attention will focus on the electronic $V$-representation problem where the tasks is to construct a KS system governed by a time-dependent potential $V^{KS}(t)$ such that the KS density matches the density of a specified interacting many-electron system at all times. We refer to this as the $V^{KS}$-representation problem.

The $V^{KS}$-representation problem associated with the model KS system is the task of constructing the potential for a model system such that the non-interacting model attains a target density expectation value. This task is not always possible and questions of $V$-representability refer to the existence of solutions to the $V^{KS}$-representation problem when different constraints are placed on the KS or on the target density.  Here, attention will primarily focus on the time-dependent $V^{KS}$-representation problem where the task is to construct a time-dependent potential $V^{KS}(t)$ such that the KS density matches a given target density at all times. 


For fermionic simulations, Ref.~\cite{Leeuwen99} was first to give a constructive solution to time-dependent $V^{KS}$-representation problem. This was challenged in Ref.~\cite{Baer08} where counter-examples were presented to this construction. These counter-examples were largely addressed by Refs.~\cite{Li08,Farzanehpour12} through a detailed analysis of densities evolving on lattices.  In separate work, implicit solutions using a fixed point mapping has been formulated directly in the continuum limit using various techniques~\cite{Ruggenthaler11,Ruggenthaler12,Ramsden2012,Nielsen13,Nielsen2018,Hodgson2013,Hodgson2016}. There is also a PDE constrained method~\cite{Jensen2016}.   Here, we will present an explicit method based on the lattice algorithm analyzed in \cite{Whitfield14a}. 

The paper begins with the description of the lattice basis used, followed by the description of the force-balance equation. We then turn towards the implementation details of the solver for the $V^{KS}$-representation problem.  A novel method for preparing consistent initial KS states for one-electron systems is given. We give some numerical examples before discussing single-electron $V$-representability theorems. Finally, we draw conclusions and point to next steps as we
close the article.

\subsection{Lattice basis}
The one dimensional basis functions $s_m\left(x\right)=\braket{x}{s_m}$ that we use are defined in Ref \cite{Colbert1992} as
\begin{equation}
    \braket{x}{s_m}=\frac{\mbox{sin}\left[\frac{\pi\left(x-\mdx\right)}{\dx}\right]}{\pi\left(x-\mdx \right)}
\end{equation}
Here the functions are  defined on domain $-\infty<x<\infty$, $m=0,\pm 1,\pm 2, ...$.  Each of the basis functions possess the property that 
\begin{equation}
   \braket{x=m'\dx}{s_m}=\frac{1}{\dx}
    \delta_{m,m\p}
\end{equation}
If infinitely many basis functions are used, the kinetic energy operator matrix $\bs{T}$ is given exactly as 
\begin{eqnarray}
    \bs{T}_{m m\p}&=&\left\{
    \begin{array}{lc}
    \frac{-\hbar^2}{2 m \dx^2}
    \frac{\pi^2 }{3},&m=m\p\\
    \frac{-\hbar^2}{ m \dx^2}
    \frac{\left(-1\right)^{m-m'}}{\left(m-m\p\right)^2},&m\neq m\p
    \end{array}
    \right.
\end{eqnarray}
 which is equivalent to an infinite order finite difference approximation \cite{Colbert1992}.  
The potential energy matrix ($\bs{V}$) in this basis is
\begin{equation}
    \bs{V}_{m m\p}=\hat{V}\left(m\dx\right)\delta_{m,m\p}=\diag(\bs{v}_m)
\end{equation}
where $\hat{V}\left(x\right)$ is the potential operator for the given system and $\bs{v}$ is the vector of diagonal elements of $\bs{V}$.

In practice, the range of $m$ is restricted to a finite number which introduces errors to $\bs{T}$ and $\bs{V}$ but contribute minimally if $m$ extends over the range where the wavefunction is appreciable. The total number of one-dimensional grid points we use is given by $L$ and is always an odd integer such that $m=0,\pm 1,\pm 2,...,\pm \frac{L-1}{2}$.

In multiple dimensions, a product basis is used such that $s_{\vec{m}}\left(\vec{x}\right)=s_{m_1}\left(x_1\right)s_{m_2}\left(x_2\right)...s_{m_D}\left(x_D\right)$, 
where $\vec{m}=\left[m_1,m_2,...,m_D\right]$, 
$\vec{x}=x_1,x_2,...,x_D$. The total number of basis functions is $L^D$. The kinetic energy operator matrix is then given by
\begin{equation}
  {T}_{\vec{m},\vec{m}\p}=\sum_i^D {T}_{m_i,m_i\p}\prod_{j\neq i}^D\delta_{m_j,m_j\p}
\end{equation}
The potential energy matrix remains diagonal with elements $\bs{V}_{\vec{m}\vec{m}}=\hat{V}\left(m_1\dx,m_2\dx,...m_D\dx \right)=\bs{v}_{\vec{m}}$. In practice, this product basis will only be viable up to around four or five dimensions. For more dimensions, it will be necessary to prune the indices $\vec{m}$.
The scaling of matrix-vector products in this basis is $\mathcal{O}\left(D L^{D+1}\right)$ for the kinetic energy operator and $\mathcal{O}\left(L^D\right)$ for the potential (as it is diagonal)\cite{Larsson2016}.

Before discussing the force-balance equation in the next section, we must introduce a few notations and definitions.  First, we will denote the KS system's wave functions as $\Phi_t$.  The one-particle wave functions are defined via an expansion over the basis functions listed above as
\begin{equation}
    \braket{x}{\phi^{(i)}_t}=\sum_m ^{(i)}{c}_m \braket{x}{s_m}
\end{equation}

Corresponding to each basis function $s_m$, there exists a creation operator $\hat a_m$ with the property the $\hat a_m\hat a_j^\dag+\hat a_j^\dag \hat a_m=\delta_{mj}$ and $\hat a_j \hat a_i+\hat a_i \hat a_j=0$.  These anti-commutation relations ensure that a wave function given by $\prod_j \left(\hat a_j^\dag \right)^{k_j}\ket{vac}$ is antisymmetric when $\ket{vac}$ is the state with no particles occupied and $k_j\in \{0,1\}$.



\subsection{The force-balance equation}
The non-interacting $V^{KS}$-representation problem requires that the fictitious system's wave function, $\ket{\Phi_t}$, evolve such that its density expectation value, $\bra{\Phi_t}\hat {\bf n}\ket{\Phi_t}$, matches a target evolution, ${\bf n}^{aim}(t)$. Here, and throughout, $\hat {\bf  n}$ is the  operator-vector with components $\hat n_j=\hat a_j^\dag \hat a_j$.
The force-balance equation determines the required instantaneous potential to correctly construct the KS system.  Note that forces enter at second order of evolution as anticipated by Newton's law: $\bos{F}=m\bos{a}$.

The force-balance equation is easily derived from the second derivative of the density following the  Heisenberg equation~\cite{Whitfield14a,Farzanehpour12}.
If we aim for a target evolution, then we should have that $\partial_t^2{\bf n}^{aim}$ (from the target interacting system) is equal to
$i\bra{\Phi_t}  [\hat {\bf H}, \partial_t \hat {\bf n}]\ket{\Phi_t}$ (from the KS system). Expanding this commutator, we have two terms, $i[\hat {\bf T},\partial_t \hat {\bf n}]$ and $i[\hat {\bf V},\partial_t \hat {\bf n}]$ which we will physically interpret as well as giving some guidance on numerical implementation.

We first discuss the acceleration which the forces must cause.
The free acceleration, also called the momentum-stress tensor \cite{Leeuwen99}, $\hat q_x=i[ \mathbf{\hat T},\partial_t \hat n_x]=-[\hat {\bf T},[\hat {\bf T},\hat n_x]]$ 
is independent of the potential operator.  
To evaluate the free acceleration, we must evaluate the expression $\hat {\bf q}=2\Re [\textrm{diag}\left(\hat {\bf T}\bos{\hat \rho} \hat {\bf T}+\bos{\hat \rho} \hat {\bf T}^2 \right)]$ evaluated against state $\Phi_t$. This can be evaluated efficiently using matrix-vector products. The expectation value of the operator $\rho$ is
\begin{equation}\label{Eq.rhoks}
{\rho}^{\Phi_t}_{jk}=\bra{\Phi_t}\hat a_{j}^{\dag}\hat a_k\ket{\Phi_t}=\sum_i^{N}    \, ^{(i)}\bs{c}_j    ^{(i)}\bs{c}_k^{\dagger}\end{equation}
where $N$ is the number of non-interacting electrons in the Kohn-Sham system. Evaluating $\bf q$ for each particle before obtaining the real part is
\begin{equation}
q_{\vec{m}}=2 \Re[\sum_{i}^{N}  \,^{(i)}\bs{c}_{\vec{m}}\p\,    ^{(i)}\bs{\bar{c}}_{\vec{m}}\p-    \,^{(i)}\bs{c}_{\vec{m}}    \,^{(i)}\bs{\bar{c}}_{\vec{m}}^{\prime \prime}]
\end{equation}
where
\begin{equation}
    \begin{array}{l}
    \,^{(i)}\bs{c}\p=\bs{T}\;^{(i)}\bs{c}\\
    \,^{(i)}\bs{ c}^{\prime \prime}=\bs{T}\;^{(i)}\bs{c}\p
    \end{array}
\end{equation}
and $\bar{ \bs{c}}$ signifies the complex conjugate of $\bs{c}$. Calculating $\bf q$ has a total scaling cost of $\mathcal{O}\left(2N DL^{D+1}\right)$.

The forced acceleration is given by the difference between the free acceleration and the target acceleration ${\bf s}=\partial_t^2 {\bf n}^{aim}-{\bf q}$.  The forced acceleration then determines the forces required from the potential.  

The forces enter through the term $i[\hat V,\partial_t \hat {\bf n}]$ which can be recast into two useful forms; one illustrating a connection to forces and the other geared towards determining the potential. 
The first form we examine is 
\begin{equation}
i[\hat V,\partial_t \hat { n}_j]=
\sum_k({v}_j-{v}_k) {T}_{kj}(  \hat{a}_j^\dag  \hat{a}_k+ \hat{a}_k^\dag  \hat{a}_j ).
\end{equation}
This form gives a nice analogy to the real space forces as $F(x)=-\nabla V(x)$.
We note that $i\bra{\Phi_t}[\hat V,\partial_t \hat  n_j]\ket{\Phi_t}=2\sum_k({v}_j-{v}_k) \; \Re [{T}_{kj}^{\Phi_t}] $ has the form of a generalized discrete gradient.  Here, the real part of ${T}^{\Phi_t}_{jk}= {T}_{jk}\bra{\Phi_t} \hat  a_j^{\dag}  \hat a_k\ket{\Phi_t}$ includes both the influence of the probability mass at each site as well as the underlying spatial metric.

The second form is more applicable to numerical simulation: 
$i\bra{\Phi_t}[\hat V,\partial_t \hat n_j]\ket{\Phi_t}$ $=\sum_r\bra{\Phi_t}\left(i[\hat n_r,\partial_t \hat n_j]\right)\ket{\Phi_t} {v}_r$.
To introduce a simpler expression, it will be advantageous to define $\mathcal{M}(\bos{A})_{ij} = {A}_{ij}-\delta_{ij}\left( \sum_k {A}_{jk}\right)$.  Then, the force-balance equation can be expressed as:
\begin{equation}
i\bra{\Phi_t}[\hat V,\partial_t \hat n_r]\ket{\Phi_t}=\sum_s\left(-2\mathcal{M}(\Re[\bs{T}^{\Phi}])\right)_{rs} {v}_s
\end{equation}
For consistency with our publication \cite{Whitfield14a}, we define $\bs{K}=-2\mathcal{M}\left(\Re[\bs{T}^{\Phi}]\right)$ as the force-balance operator.  For symmetric matrices, $\bs{A}=\bs{A}^{T}$, $\mathcal{M}(\bs{A})$ will have the constant vector in its null space.  The gauge freedom physically stems from the irrelevance of the zero of energy. Since we are concerned with time-dependent quantum mechanics, the constant potential only imprints an unobservable global phase on the wave function.  

\section{$V^{KS}$-representation solver}
Elsewhere \cite{Whitfield14a}, we presented and analyzed an explicit solution for the time-dependent potential necessary for TDDFT provided with the density time-trace of a $V$-representable system.  The algorithm was found to scale polynomially in all input parameters except for the $V$-representability parameter \cite{Whitfield14a} which diverges exponentially when the interacting system no longer has a corresponding KS system.  

The algorithm requires, as inputs, the complete time-trace of the density and a consistent initial state $\Phi$ that reproduces the initial density and the initial time derivative of the density.  For numerical implementation, a representation of the kinetic operator in the lattice basis is also needed.  

\subsection{Preparing consistent initial KS states}
\subsubsection{One-electron}
Unlike previous numerical studies~\cite{Li08,Baer08,Verdozzi08,Ruggenthaler12,Schmitteckert13,Nielsen13,Fuks14}, we consider preparation of initial KS states with non-zero initial momentum.
Suppose that the initial KS system should yield an initial time derivative given by $\dot{\bs{n}}^{aim}$.  
Using the Heisenberg equation of motion it can be shown \cite{Whitfield14a}, that for a one-electron wave function in a lattice basis, $\psi$, the density derivative is given by $\dot n_j=-i \sum_k {T}_{ki} (\langle \hat a_i^\dag \hat a_k \rangle_\psi-\langle \hat a_k^\dag \hat a_i \rangle_\psi)$.  Given that we consider a single electron wave function, the 1-RDM elements can be written as $\langle \hat a_i^\dag \hat a_j\rangle_{\psi}=\sqrt{n_in_j}e^{i(\theta_j-\theta_i)}$. 

To assign the phases, roughly speaking, we must solve the equation: $\Delta\boldsymbol{\theta}=\frac{\partial\boldsymbol{\theta}}{\partial \dot{ \bs{n}}} \Delta \dot{\bs{n}}$.  This is the content of Newton's method.  We will describe the modifications needed to handle the gauge degree of freedom after deriving the Jacobian, $J_{ij}=\frac{\partial \dot n_i}{\partial\theta_j}$.

The Jacobian is also given by (minus) the force-balance matrix:
\begin{eqnarray}
\frac{\partial \dot n_i}{\partial \theta_j}
&=&\frac{\partial}{\partial\theta_j}\left(-i \sum_k {T}_{ki} (\langle \hat a_i^\dag \hat a_k \rangle_\psi-\langle  a_k^\dag a _i \rangle_\psi)\right)\nonumber\\
&=&2\sum_k {T}_{ki} \sqrt{n_kn_i}\; \left[\frac{\partial\sin(\theta_k-\theta_i)}{\partial\theta_j}\right]\nonumber\\
&=&2\sum_k {T}_{ki}\sqrt{n_kn_i}\cos(\theta_k-\theta_i)\left[
 \delta_{jk}
-\delta_{ij}
\right]\nonumber\\
&=& {T}_{ij}(\langle\hat a_i^\dag\hat a_j\rangle_\psi+\langle\hat a_j^\dag \hat a_i\rangle_\psi)\nonumber\\
&&-\delta_{ij}\sum_k
{T}_{jk}(\langle\hat a_k^\dag \hat a_j\rangle_\psi+\langle \hat a_j^\dag \hat a_k\rangle_\psi)\nonumber\\
J_{ij}=\frac{\partial \dot n_i}{\partial \theta_j}&=&-{K}_{ij}
\end{eqnarray}

Before applying the Newton method, we must account for the gauge corresponding to the global phase of the wave function. Other manifestations of this gauge degree of freedom are 1) the one-dimensional null space of the Jacobian and 2) the constraint that $\sum \dot n_j=0$.

We can fix the phase of one of the $M$ wave function components in order to fix the gauge.  Suppose the fixed phase is the component with the largest $n_m$ at position $p$ (which we assume is the best described lattice point), then we only update $\vec\theta_g$ on the $M-1$ remaining components.  We have found this to be superior to just choosing an arbitrary component to fix the phase.

Putting it all together, the Newton rule for updating the phase vector $\vec \theta$ to a new assignment $\boldsymbol\vartheta$ is
\begin{equation}
\boldsymbol{\vartheta}=\boldsymbol{\theta} -(\Delta \boldsymbol{\theta}- \Delta \boldsymbol{\theta}_p)
\end{equation}
where component $\Delta\boldsymbol{\theta}_p$ is subtracted so that the $p$th component is not updated.  We have defined
\begin{equation}
    \Delta \boldsymbol{\theta}=\bs{J}^{-1} (\dot{\bs{n}} -\dot{\bs{n}}^{aim}) =-\bs{K}^{-1}(\dot{\bs{n}} -\dot{\bs{n}}^{aim})
\end{equation}
Note that dropping the gauge component to get $\dot n_g$ and $\dot n^{aim}_g$ loses no information due to the constraint that $\sum \dot n_j=0$.

Numerical results with a straightforward implementation of Newton's method works quite well provided that the wave function's initial momentum is somewhat close to the target momentum. 

\subsubsection{Multi-electron systems}
For $N$ non-interacting one-electron wave functions in a lattice basis with wavefunctions $\phi^{(i)}$, the total density derivative is given by $\dot n_j=-i\sum_{i}^{N} \sum_k {T}_{kj} (\langle a_i^\dag a_k \rangle_{\phi^{(i)}}-\langle  a_k^\dag  a_j \rangle_{\phi^{(i)}})$.  The 1-RDM elements can be written as $\langle a_j^{\dag} a_k \rangle_{\Phi}=\sum_{p}^{N}\sqrt{^{(p)}n_j \,^{(p)}n_k}e^{\iu (\theta_k^{(p)}-\theta_j^{(p)})}$. 

Following the section above, the Jacobian is given as
\begin{equation}
\frac{\partial \dot n_j}{\partial\, \theta_k^{(i)}} =-{K}^{(i)}_{jk},
\end{equation}
where $\bs{K}^{(i)}$ denotes the force-balance matrix that only includes the i-th one-electron wavefunction of the $N$ non-interacting wavefunctions. To obtain the next approximation of phases, we need to solve the under-determined least squares problem
\begin{equation}
    \bs{K}^{\Phi_t}\Delta  {\boldsymbol{\theta}}^{\Phi_t}=\dot {\bf n}-\dot {\bf n}^{aim},
\end{equation}
where $\bs{K}^{\Phi_t}=\left[\bs{K}^{(1)},\bs{K}^{(2)},...,\bs{K}^{(N)}\right]$ and $\Delta \boldsymbol{\theta}^{\Phi_t}=\left[\Delta \boldsymbol{\theta}^{(1)},\Delta \boldsymbol{\theta}^{(2)},...,\Delta \boldsymbol{\theta}^{(N)}\right]^T$. In this paper, the solution is obtained iteratively using a least squares method as implemented by \Ref{SOL} or a Morse-Penrose pseudo inverse. In two and three dimensions, the iterative method will likely be preferred.  The Newton steps taken for each particle are then
\begin{equation}
\boldsymbol{\vartheta}^{(i)}=\boldsymbol{\theta}^{(i)} -(\Delta \boldsymbol{\theta}^{(i)}- \Delta \boldsymbol{\theta}_p^{(i)}).
\end{equation}
Here $\Delta \boldsymbol{\theta}_p$ is the zero vector with the $p$th entry equal to $\Delta\theta_p$. Note that the gauge needs to be fixed individually for all $N$ electrons.

\subsection{Numerical inversion of the force-balance matrix}
The inversion of the force-balance matrix has a few problems. It always includes a null space of the constant vector and can be poorly conditioned when parts of the density are very small. If the KS system does not have a single potential corresponding to a given momentum-stress tensor, then other singular vectors can be introduced. 

In order to exploit the structure of the force-balance matrix, and the corresponding $\hat{\rho}$, we use iterative Krylov-space methods to solve for $\bs{K}\bs{b}=\bs{x}$. The 1RDM  $\boldsymbol{\rho}$ is a rank-$N$ matrix with matrix elements given in \Eq{Eq.rhoks}. The force-balance matrix is
\begin{equation}\label{Eq.fbm}
    \bs{K} =-2\Re[ \bs{T} \circ \boldsymbol{\rho}^{\Phi} -\diag\left( \sum_k \left(\bs{T} \circ \boldsymbol{\rho}^{\Phi}\right)_{jk}\right)]
\end{equation}
where $\diag(\bs{v})$ is a diagonal matrix with elements $v_j$ as the $j$th entry and $\circ$ denoting the Hadamard product.

Using a property of Hadamard products of matrices, the first term of \Eq{Eq.fbm} can be evaluated as 
\begin{equation}\label{Eq.fbe1a}
    -2\Re[\sum_{i=1}^{N}\diag(\,^{(i)}\bs{c})\bs{T}\diag(\,^{(i)}\bs{\bar{c}})]
\end{equation}
To obtain the real part only, we note that 
\begin{equation}\label{Eq.rdc}
\bs{c}\bs{c}^{\dagger}=\bs{r}\bs{r}^T+\iu \bs{i}\bs{r}^T+
\bs{i}\bs{i}^T-\iu \bs{r}\bs{i}^T
\end{equation}
where $\bs{r}$ and $\bs{i}$ are real vectors such that $\bs{c}=\bs{r}+\iu \bs{i}$. This means that the complex rank-1 matrix $\bs{c}\bs{c}^{\dagger}$ is equivalent to the sum of two real rank-1 matrices and two imaginary rank-1 matrices. By retaining only the real matrices, \Eq{Eq.fbe1a} can be rewritten as
\begin{equation}\label{Eq.fbe1b}
    \sum_{i=1}^{N}-2\left(\diag(\,^{(i)}\bs{r})\bs{T}\diag(\,^{(i)}\bs{r})+\diag(\,^{(i)}\bs{i})\bs{T}\diag(\,^{(i)}\bs{i})\right)
\end{equation}
This term can be applied to a vector in $\mathcal{O}\left(2N (L^D+DL^{D+1}+L^{D}\right)$ by applying the matrices sequentially.\cite{Brown2018}

For the second term in \Eq{Eq.fbm}, it is best to obtain the diagonal elements of the matrix once. This is done using the property of Hadamard products\cite{Styan1973}
\begin{equation}
\sum_j\left(\bs{A}\circ \bs{B}\right)_{jk}=\left(AB^T \right)_{jj} 
\end{equation}
This means that the second term can be evaluated as
\begin{equation}
-2\Re[-\diag\left( \sum_k \left(\bs{T} \circ \boldsymbol{\rho}^{\Phi} \right)_{jk}\right)]=\diag(2\Re[\sum_{i}^{N}\,^{(i)}\bs{D}])
\end{equation}
where
\begin{equation}
   ^{(i)}D_{jk}=\,^{(i)}\bs{c}_j\p  \,^{(i)}\bar{\bs{c}}_k 
\end{equation}
where as before, $\,^{(i)}\bs{c}\p=\bs{T}\,^{(i)}\bs{c}$,
and $ ^{(i)}\bar{\bs{c}}$ signifies the complex conjugate of $^{(i)}\bs{c}$. 

In one dimension, there is no advantage to performing the matrix-vector products this way. However, in multiple dimensions, the scaling of the matrix-vector products is much better and the storage requirements for the matrix are also better with order $\mathcal{O}(L+(2N+1)L^D)$ instead of $L^{2D}$.

\subsection{Iterative solver}
In order to solve for the potential, we use the MINRES-QLP algorithm\cite{Choi2011}. This method is designed to solve singular Hermitian problems like that of $\bs{K}\bs{v}=\partial_t^2{\bf n}^{aim}-\bf{q}$. 

For obtaining the initial state phases, we use the LSQR algorithm\cite{Paige1982}. This  algorithm  requires the application of $\bs{K^{\Phi}} \bf x$ and $\bs{K^{\Phi}}^T\bf y$ to solve the matrix problem $\bs{K^{\Phi}}\bf x=y$. This poses no problem when exploiting the structure of the force-balance matrix. Applying  $\bs{K^{\Phi}}$ involves applying each $\bs{K}^{(i)}$ to the corresponding segment of $\bf x$ while  $\bs{K^{\Phi}}^T\bf y$ involves applying each $\bs{K}^{(i)}$ to $\bs{y}$ in succession. 

\subsubsection{Preconditioning}
Without preconditioning, the force-balance matrix is poorly conditioned due to the large changes in the density on the grid. In our testing, it appears that using a diagonal preconditioner works quite well. This is defined as
\begin{equation}
    M_{\vec{m}\vec{m}\p}=
    \frac{1}{\left|K_{\vec{m}\vec{m}}\right|}\delta_{\vec{m}\vec{m}\p}
\end{equation}
where the new problem is
\begin{equation}\label{Eq.Vsolve}
     \bs{M}^{1/2}\bs{K}\bs{M}^{1/2}(\bs{M}^{-1/2}\bs{v})=\bs{M}^{1/2}\bs{s}
\end{equation}
For the assigning of phases, the other half of the preconditioning matrices is 
\begin{equation}
    \bs{M^{\Phi}}=
    \left[\begin{array}{cccc}
         \bs{M^{(1)}}&0&0&0  \\
         0& \bs{M^{(2)}}& 0 &0\\
         0& 0&\ddots &0\\
         0& 0& 0& \bs{M^{(N)}}
    \end{array}\right]
\end{equation}
where
\begin{equation}
    M^{(i)}_{\vec{m}\vec{m}\p}=
    \frac{1}{\left|{K^{(i)}_{\vec{m}\vec{m}}}\right|}\delta_{\vec{m}\vec{m}\p}
\end{equation}
with the problem now,
\begin{equation}
    \bs{M}^{1/2}\bs{K^{\Phi}}\bs{M^{\Phi}}^{1/2}(\bs{M^{\Phi}}^{-1/2}\bs{\Delta\boldsymbol{\theta}^{\Phi}})=\bs{M}^{1/2}\left(\dot {\bf n} -\dot {\bf n}^{aim}\right).
\end{equation}

If the value of $K_{\vec{m},\vec{m}}$ is zero (or $K^{(i)}_{\vec{m},\vec{m}}$) drops below $10^{-15}\times \mbox{max}\left(\bs{K}\right)$  (or $10^{-15}\times \mbox{max}\left(\bs{K}^{(i)}\right)$), then $M_{\vec{m},\vec{m}}$ is set to zero. The density at lattice point $\vec{m}$ is then deemed too small to resolve a potential and is removed from consideration. As both MINRES-QLP and LSQR are for singular problems, this poses no difficulties. In the future, it would be more efficient to prune those points from the required matrix-vector products to improve efficiency \cite{Brown2018,Larsson2016}.

\subsection{Time Propagator}\label{sec.timeprop}
The method we use to solve the TDSE is based on \Ref{vanDijk2014}. The Schr\"{o}dinger equation that is solved is defined as
\begin{equation}\label{Eq.vdH}
    \frac{\partial\bos{\psi}\lrb{t}}{\partial t}=-\frac{\iu}{\hbar}\lrb{\bs{H_0}+\tilde{\bs{V}}\lrb{t}}\bos{\psi}\lrb{t}
\end{equation}
where $\bs{H_0}$ is the time-independent portion of the Hamiltonian (that we assume is given) and $\tilde{\bs{V}}\lrb{t}$ is the time-dependent potential (which is the goal) at time $t$. $\bs{H_0}$ only includes $\bs{T}$, but can also include
the external one-electron potential if it is given as part of the problem instance and is time-independent. 

This method propagates wavefunctions using,
\begin{equation}\label{Eq.vdprop}
    \bos{\psi}\left(t\right)=\frac{e^{-\iu \bs{H_0} \Delta t}\bos{\psi}
    ^{(-)}\lrb{t-\Delta t}+\bos{f}\lrb{t:\bos{\psi}\lrb{t}}}{\bs{I}+\frac{\iu \Delta t}{2\hbar}\tilde{\bs{V}}\lrb{t}}
\end{equation}
where
\begin{equation}
    \bos{\psi}^{(-)}\lrb{t}=\lrb{\bs{I}-\frac{\iu \Delta t}{2\hbar}\tilde{\bs{V}}\lrb{t-\Delta t}}\bos{\psi}\lrb{t-\Delta t}-\bos{f}\lrb{t:\bos{\psi}\lrb{t-\Delta t}}
\end{equation}
We only retain the next highest
Euler-MacLaurin expansion term in $\bos{f}\lrb{t:\bos{\psi}\lrb{t}}$ which for our purposes is defined as, 
\begin{equation}
    \bos{f}\lrb{t:\bos{\psi}\lrb{t}}=\frac{\Delta t}{12}\lrb{-\bs{H_0}\tilde{\bs{V}}\lrb{t}\bos{\psi}\lrb{t}+
    \iu \lrb{\frac{\partial \tilde{\bs{V}}\lrb{t}}{\partial t}\bos{\psi}\lrb{t}+\tilde{\bs{V}}\lrb{t}\frac{\partial\bos{\psi}\lrb{t}}{\partial t}}}
\end{equation}
with \Eq{Eq.vdH} defining $\frac{\partial\bos{\psi}\lrb{t}}{\partial t}$ and the method to obtain $\frac{\partial \tilde{\bs{V}}\lrb{t}}{\partial t}$ described later.

\Eq{Eq.vdprop} is an implicit equation in that ${\bos f}\lrb{t:\bos{\psi}\lrb{t}}$ depends on $\bos{\psi}\lrb{t}$. Here, both ${\bos f}\lrb{t:\bos{\psi}\lrb{t}}$ \emph{and} $\tilde{\bs{V}}\lrb{t}$  are defined by $\bos{\psi}\lrb{t}$. Therefore the algorithm to update the potential and the wavefunction is an iterative process where,
\begin{eqnarray}\label{Eq.iterprop}
\bos{\psi}^{(k+1)}\left(t\right)=\frac{\bos{\psi}^{(+)}\lrb{t}+{\bos{f}}\lrb{t:\bos{\psi}^{(k)}\lrb{t}}}{\bs{I}+\frac{\iu \Delta t}{2\hbar}{\bs{\tilde{V}}^{(k)}(t)}}
\end{eqnarray}
where $\bos{\psi}^{(+)}\lrb{t}=e^{-\iu \bs{H_0} \Delta t}\bos{\psi}^{(-)}\lrb{t-\Delta t}$ is calculated once using the short iterative Arnoldi propagator. Initially, ${\tilde{\bs{V}}^{(0)}\lrb{t}}=\tilde{\bs{V}}(t-\Delta t)$ and $\bos{\psi}^{(0)}\left(t\right)=\bos{\psi}\left(t-\Delta t\right)$. We first update only
${\bos{f}}\lrb{t:\bos{\psi}^{(i)}\lrb{t}}$ while keeping ${\bs{\tilde{V}}^{(k)}(t)}$ unchanged and iterate \Eq{Eq.iterprop} until $\left|\bos{\psi}^{(k+1)}-\bos{\psi}^{(k)}\right|$ is below machine precision. We then update $\bs{V}^{(t)}$ using \Eq{Eq.Vsolve} and update ${\bos f}\lrb{t:\bos{\psi}^{(i)}\lrb{t}}$ followed by the iteration of \Eq{Eq.iterprop}.  This process is repeated until the difference in $\bos{\psi}\lrb{t}$ before and after updating $\bs{V}\lrb{t}$ is below some threshold. Note that the time-independent external potential will need to be subtracted from the result of \Eq{Eq.Vsolve} if it was given as part of the problem instance specification.

\subsection{Obtaining the partial derivative of the potential}
The definition of ${\bos f}\lrb{t:\bos{\psi}^{(i)}\lrb{t}}$ includes a partial derivative of $\tilde{\bs{V}}(t)$ with respect to time. One can introduce an error in \Eq{Eq.iterprop} by assuming that $\frac{\partial \tilde{\bs{V}}(t)}{\partial t}=0$. However, we use a finite difference approximation to $\frac{\partial \tilde{\bs{V}}(t)}{\partial t}$ which improves the accuracy of the results. 

We initially define the zero of energy by some value $E_0$, which is usually $E_0=0$. Each time a potential $\tilde{\bs{V}}(t)$ is obtained, we shift the result such that the total energy of the system remains at $E_0$. This total energy is defined as
\begin{equation}
    E=\sum_{i=1}^{N}\,^{(i)}\bs{c}^{\dagger}\lrb{\bs{H_0}+\tilde{\bs{V}}(t)}\,^{(i)}\bs{c}
\end{equation}
and results in the correctly shifted potential of $\tilde{\bs{V}}(t)\rightarrow \tilde{\bs{V}}(t)-(E-E_0)/N$. This method appears to correctly position the potential for the systems examined. The method would need to be modified if the multi-electron system has a source term imparting energy on the system. 

To obtain the derivative of the potential, we use the backward finite-difference approximation. Therefore, $\frac{\partial \tilde{\bs{V}}(t=0)}{\partial t}=0$. $\frac{\partial\tilde{\bs{V}}(t=\Delta t)}{\partial t}=(\tilde{\bs{V}}(t)-\tilde{\bs{V}}(t-\Delta t))/\Delta t$, and $\frac{\partial\tilde{\bs{V}}(t=j\Delta t)}{\partial t}=(\tilde{\bs{V}}(t)-4\tilde{\bs{V}}(t-\Delta t)+\tilde{\bs{V}}(t-2\Delta t))/(2\Delta t)$ for $j>1$. Higher order finite-difference approximations did not improve results.

The method and techniques outlined above generally behave well in regions where the density is appreciable. However, it is helpful for stability to assign the phases after every time step which requires that the first derivative of the density is provided. This should not be a problem as the second derivative of the density is required for potential inversion. The first derivative can also be obtained by measuring the full one-body reduced density matrix. The first derivative of the density is then given as $i\partial_t \hat{\bos{\rho}}=\left[\hat{\bos{T}},\hat{\bos{\rho}}\right]$. 

One stability issue that can occur is very large changes in the inverted potential where the density is small due to numerical artifacts. This can be attenuated by setting a maximum allowed change in the potential for each lattice point at each time step.

\section{One-electron test}
The one-electron test we examine is a sum of two coherent wavepackets in a Harmonic potential. The potential is constant at all times as
\begin{equation}
    V(x,t)=\frac{1}{2}\omega x^2,
\end{equation}
where $\omega=1$.
The wavefunction initial state is
\begin{equation}
    \Psi\lrb{x,t=0}=Ae^{-x^2/2}\cos{\left[\frac{5\sqrt{\pi}}{2}x\right]}
\end{equation}
where $A$ is a normalization factor. This is the superposition of the two coherent states with momentum of $\pm 5\sqrt{\pi}/2$. The grid used has 115 points and covers a range of $x\in \left[-11,11\right]$ with 1600 time steps to obtain a total time of $T/2$ where $T=2\pi/\omega$. The potential inversion is performed using a pseudoinverse of \Eq{Eq.Vsolve} instead of MINRES-QLP. For one-dimension, there is no advantage to using MINRES-QLP.
\begin{figure}
\begin{subfigure}{.5\textwidth}
  \centering
  \includegraphics[width=.8\linewidth]{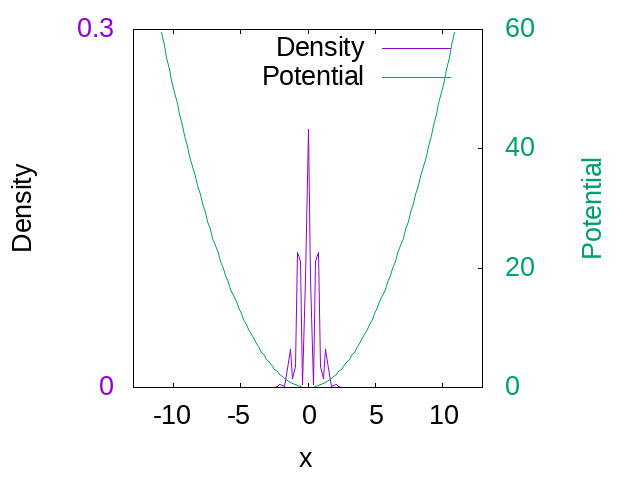}
  \caption{Initial density and potential.}
  \label{fig:sfig1a}
\end{subfigure}%
\begin{subfigure}{.5\textwidth}
  \centering
  \includegraphics[width=.8\linewidth]{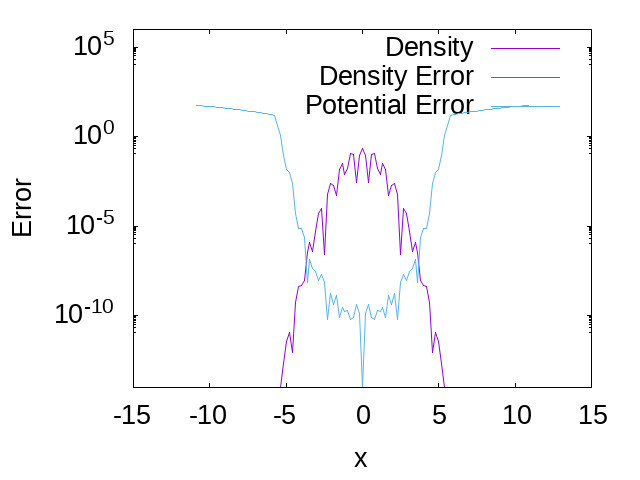}
  \caption{Initial error in the potential. Density error at t=0 is defined to be zero.}
  \label{fig:sfig1b}
\end{subfigure}
\begin{subfigure}{.5\textwidth}
  \centering
  \includegraphics[width=.8\linewidth]{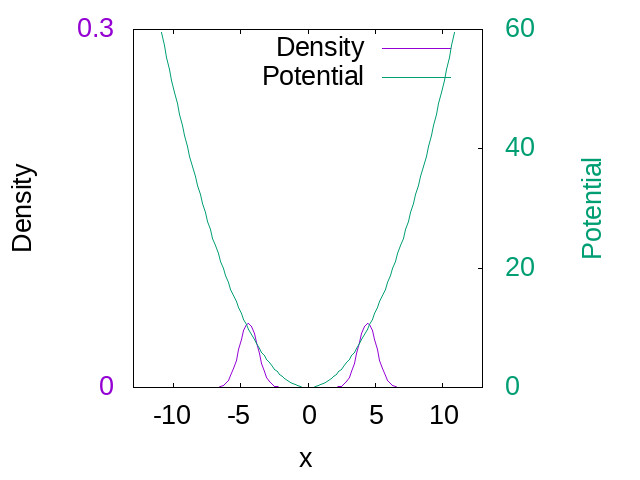}
  \caption{Exact density and potential at $t=T/4$.}
  \label{fig:sfig1c}
\end{subfigure}%
\begin{subfigure}{.5\textwidth}
  \centering
  \includegraphics[width=.8\linewidth]{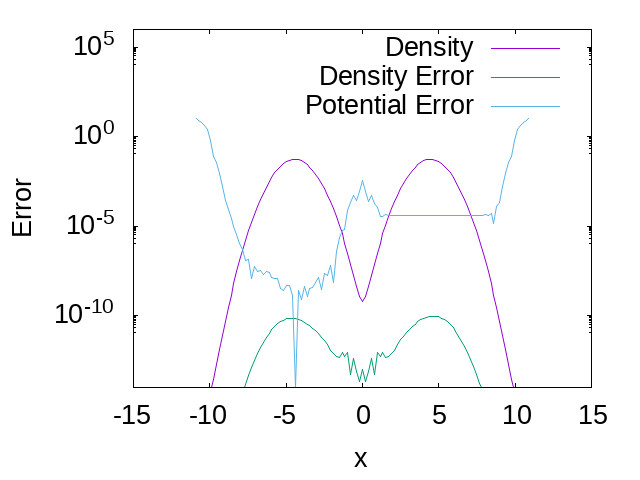}
  \caption{Density and potential error at $t=T/4$.}
  \label{fig:sfig1d}
\end{subfigure}
\begin{subfigure}{.5\textwidth}
  \centering
  \includegraphics[width=.8\linewidth]{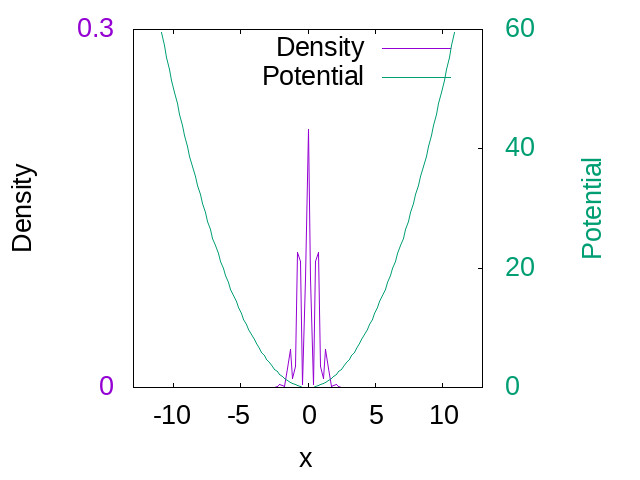}
  \caption{Exact density and potential at $t=T/2$.}
  \label{fig:sfig1e}
\end{subfigure}%
\begin{subfigure}{.5\textwidth}
  \centering
  \includegraphics[width=.8\linewidth]{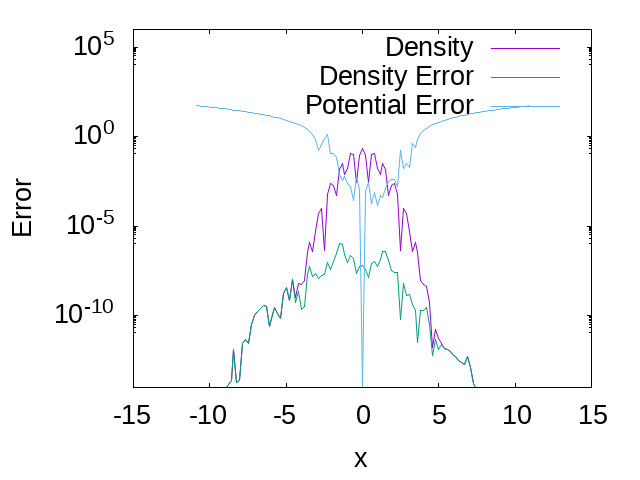}
  \caption{Density and potential error at $t=T/2$.}
  \label{fig:sfig1f}
\end{subfigure}
\caption{Plots of the density and potential for the separating wavepacket problem.  To define the potential error, we shift the recovered potential such that they coincide at the point of highest density. This manifests as the spike to $-\infty$ on the position versus log scale error in the three subplots in the right column.}
\label{fig:separating}
\end{figure}

The exact solution is obtained by propagating the initial wavepacket using the Short Iterative Arnoldi (SIA) propagator. Snapshots of the method's performance are shown in FIG. \ref{fig:separating}.  At $t=0$ the density error is zero by definition. The potential error is inversely related to the magnitude of the density, or the error is smallest where the density is largest. At $t=T/4$, the wave packet is split into two distinct regions. Once again, the potential error is small where the density is large and larger where the density is small. However, there is another feature where a constant shift error occurs on the right side of FIG . \ref{fig:sfig1d}. This is due to the inversion method not being able to resolve the middle (low density) region enough to connect the two sides with the same potential offset. At $t=T/2$, this offset error contributes to some inaccuracy in the recovered potential. The error in the recovered potential is approximated bounded by the shift error of $\approx 10^{-4}$ from $t=T/4$.  However the results are still reasonable.  If the initial state has a larger momentum than $\pm 5\sqrt{\pi}/2$, then the density connecting the two wavepackets tends to zero at $t=T/4$ and larger shift errors can occur. It would then be necessary to separate the problem into two smaller systems with different energies. The splitting of the system could be defined with the use of the preconditioner.

This one electron test shows that our method recovers the potential accurately under a variety of circumstances. It succeeds for large changes in density at individual grid points and can ``discover'' the correct potential as the density explores the physical space. It also can resolve the potential for two higher density areas with a small density in between, and works for complicated densities with non-negligible momentum.   

\section{Two-electron test}
The two electron test that we examine is that first described in \Ref{Hodgson2013}. The initial state is the first triplet state of 
\begin{equation}\label{eq.2elecH0}
    \hat{H_0}=-\frac{\partial^2}{\partial x_1^2}-\frac{\partial^2}{\partial x_2^2}+V_{d}(x_1)+V_{d}(x_2)+\frac{1}{\sqrt{\left|x_1-x_2\right|+0.1}}
\end{equation}
where $V_d(x)=\alpha x^{10}-\beta x^4$, with $\alpha=5\times 10^{-11}$ and $\beta=1.3\times 10^{-4}$. At $t\ge 0$, the system becomes 
\begin{equation}
    \hat{H}=-\frac{\partial^2}{\partial x_1^2}-\frac{\partial^2}{\partial x_2^2}+V_{ext}(x_1)+V_{ext}(x_2)+\frac{1}{\sqrt{\left|x_1-x_2\right|+0.1}}
\end{equation}
where $V_{ext}(x_i)=V_{d}(x_i)-x_i/10$ which drives the electrons to the right. The first triplet state is calculated using ARPACK to obtain the lowest energy state.  Anti-symmetry of the spatial wave function was enforced for the starting vector and after each matrix-vector product. 

The initial state for KS system was obtained by obtaining the full 1RDM of the two-electron initial triplet state and performing a singular value decomposition (SVD). The SVD of a $L \times L$ matrix obtains the optimal truncated $R$-rank representation 
\begin{equation}
    {\boldsymbol{\rho}}\approx \sum_{i=1}^{R}\sigma_i \bs{u}_i\bs{v}_i^{\dagger}
\end{equation}
where $\bs{M}=\bs{U\Sigma V^*}$, $\bs{u}_i$ is the $i$th column of $\bs{U}$, $\bs{v}_i$ is the $i$th column of $\bs{V}$ and $\sigma_i$ is the $i$th largest singular value. As the 1RDM is hermitian positive semi-definite, $\bs{v}_i=\bs{u}_i$.  These are also the eigenvectors (natural orbitals) of the 1RDM.

In the case of the 1RDM, each $\sigma_i$ is a number that represents the occupancy of the $i$th natural orbital, $\bs{ u}_i$. For the interacting system examined at $t=0$ of \Eq{eq.2elecH0}, the first two 
singular values were degenerate with occupancy of $0.999999799989864$. This signifies that the system is very close to non-interacting initially (i.e. describable by a state with $\sigma_i$ taking only values zero or one).

To obtain initial KS orbitals, an iteration scheme is implemented. Initially, the orbitals of the KS system are taken as the first two singular vector $\bs{u}_1$ and $\bs{u}_2$. Element by element, if $u_{1i}<{\bf n}^{(1)}_i \rightarrow u_{2i}=\mbox{sgn}(u_{2i})\sqrt{{\bf n}^{(1)}_i-u_{1i}^2}$. Likewise, if $u_{2i}<{\bf n}^{(1)}_i \rightarrow u_{1i}=\mbox{sgn}(u_{1i})\sqrt{{\bf n}^{(1)}_i-u_{2i}^2}$. The vectors $\bs{u}_{1}$ and $\bs{u}_2$ are orthonormalized after each sweep through the grid points. After a number of iterations, the vectors converge such that densities ${\bf n}^{(1)}$ and ${\bf n}^{(KS)}$ exactly match and $\bs{u}_{1}$ and $\bs{u}_2$ are orthonormal. Using the natural orbitals of the 1RDM for an interacting system should produce reasonable starting vectors for any number of electrons along with the iterative method described here.  

The two-electron system is propagated using the short iterative Arnoldi method using a Krylov subspace of $50$ with exchange anti-symmetry enforced throughout the calculation. The basis functions are on a $271\times 271$ grid in the range of $x \in \left[-13.5,13.5\right]$ with $\dx=0.1$ and $dt=0.005$. The KS system is propagated using the method of section \ref{sec.timeprop} with the phases assigned at every step. The inverse of $\bs{K}$ was obtained using the preconditioner $\bs{M}$ and MINRES-QLP. The inversion of $\bs{K}$ is still difficult even when using the preconditioner for this system. There is a very small ($\approx 10^{-10}$) eigenvalue of $\bs{M}^{-1/2}{\bf K}{\bf M}^{-1/2}$. All the other eigenvalues are $>10^{-3}$ aside from the  constant  vector which is known to be in the null space.  

\begin{figure}
\begin{subfigure}{.5\textwidth}
  \centering
  \includegraphics[width=.8\linewidth]{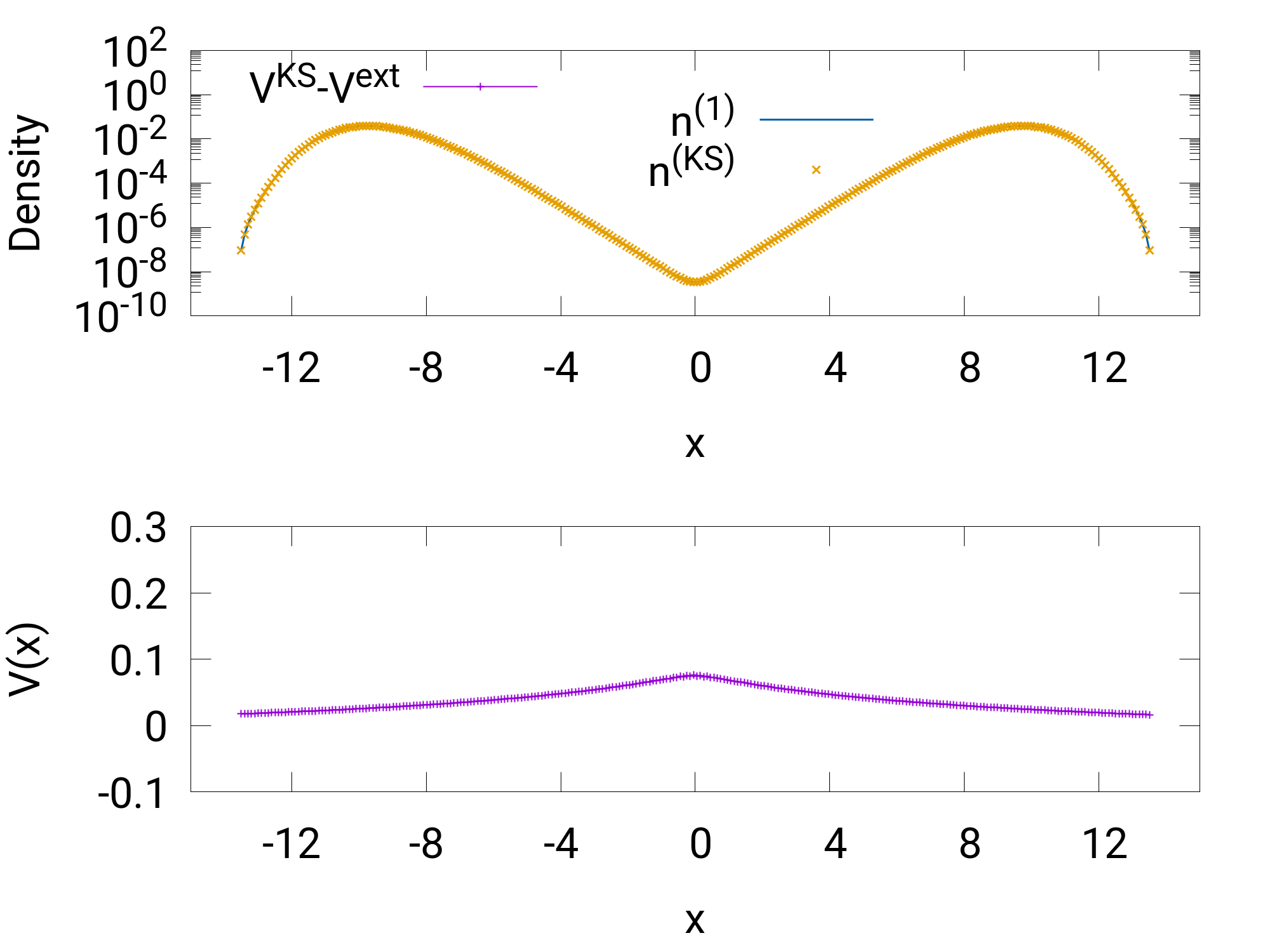}
  \caption{Initial density and potential}
  \label{fig:2sfig1a}
\end{subfigure}%
\begin{subfigure}{.5\textwidth}
  \centering
  \includegraphics[width=.8\linewidth]{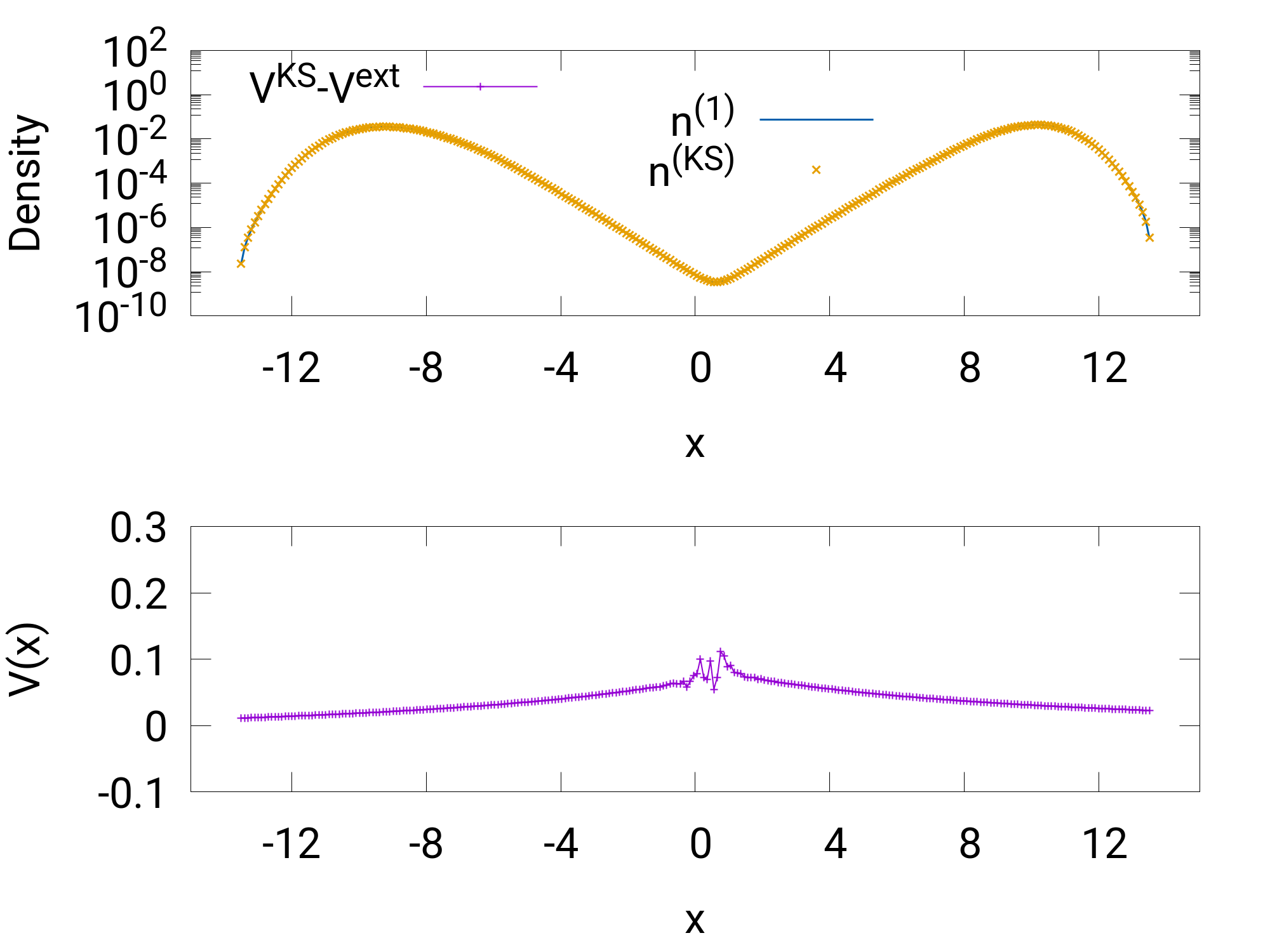}
  \caption{Density and potential at $t=3.5$}
  \label{fig:2sfig1b}
\end{subfigure}
\begin{subfigure}{.5\textwidth}
  \centering
  \includegraphics[width=.8\linewidth]{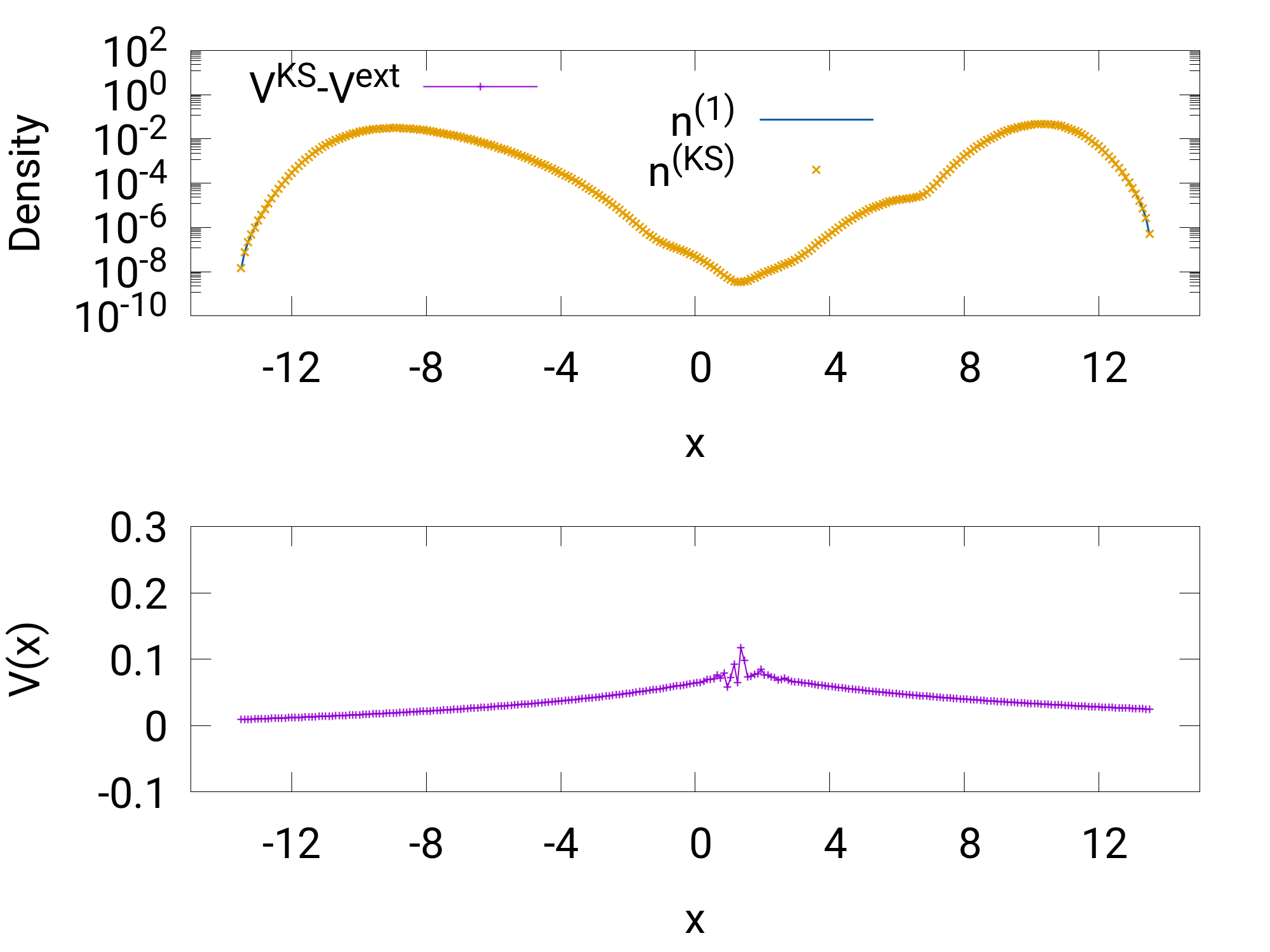}
  \caption{Exact density and potential at $t=5.3$.}
  \label{fig:2sfig1c}
\end{subfigure}%
\begin{subfigure}{.5\textwidth}
  \centering
  \includegraphics[width=.8\linewidth]{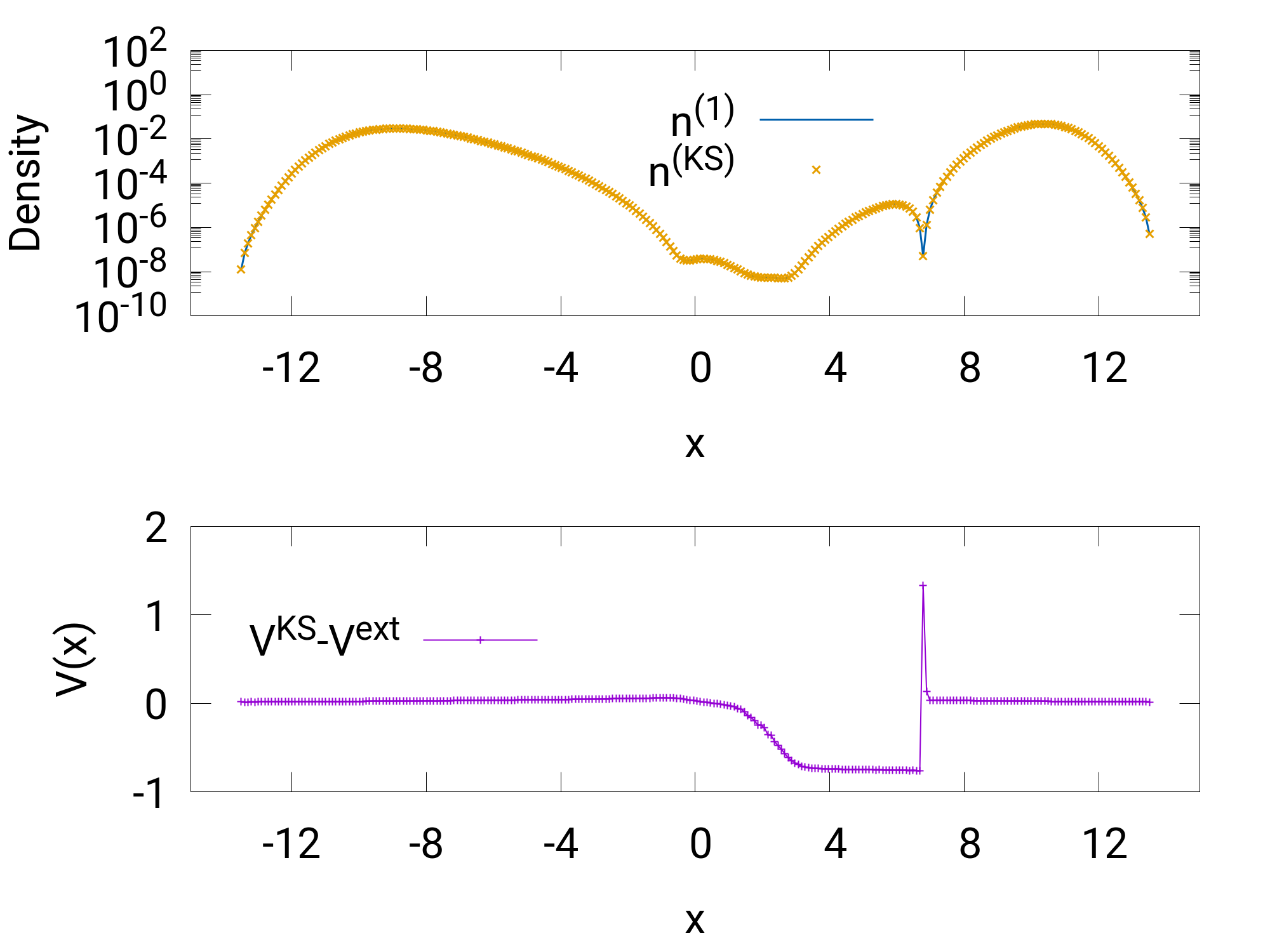}
  \caption{Density and potential at $t=5.75$.}
  \label{fig:2sfig1d}
\end{subfigure}
\caption{Plots of the log of the density and potential for the two-electron problem.}
\label{fig:2elec}
\end{figure}

The results of the potential inversion are shown in FIG. \ref{fig:2elec}. The initial $V^{KS}-V_{ext}$ is very similar to that of \Ref{Hodgson2013}. It is slightly different due to the fact that the driving potential is included in our calculation at $t=0$ and we do not use vector potentials. The times of $t=3.5$ and $t=5.3$ are included to compare with the previous studies of \Refs{Hodgson2013,Jensen2016}. \Ref{Hodgson2013} obtains  different Kohn-Sham potential in the middle region but still had the same $1/r$ decay. The primary difference is due to the fact that they used vector potentials and we have not.  \Ref{Jensen2016} could not recover a potential for this system. 

Our method fails around $t=5.75$ as shown in FIG.\ref{fig:2sfig1d}. We suspect this is due to the formation of a cusp that prevents the solution from being continued.  Shockwave formation in partial differential equations are known to cause similar issues [ref].
The cusp in the density forms at $x\approx 6.8$ starting at about $t\approx 5.70$ and appears for about 0.2 time units before disappearing.  \Ref{Hodgson2013} does not discuss this region in time so it is unclear if their method can circumvent this issues.  Despite the failure to continue the solution through the formation of this sharp feature in the density, we can easily restart the solver at time $5.9$ without further issues.



Note that we suspect that a cusp has formed since oscillatory behavior did not present a problem our method as seen in the one-electron test where the density fluctuates over its support.

\section{$V$-representability}
We now make a few comments on the existence of solutions to the $V^{KS}$-representation problem.
A previous theorem \cite{Farzanehpour12} showed that ground states of many-body interacting systems are always $V$-representable in the neighborhood of the initial time.  They showed that the matrix ${\bf K}$ has only one zero eigenvalue and is positive definite in the space of inhomogeneous potentials. It should be noted that the theorem does not characterize the $V$-representability parameter thus numerical stablility is not ensured.

Here we use simpler arguments to provide additional characterizations of the spectrum of ${\bf K}$ in the single-electron case. This coincides with the previous theorem for the ground state but generalizes to all eigenstates.

\textbf{Theorem:} Given non-degenerate $\boldsymbol{\psi}$ such that ${\bf H}\boldsymbol{\psi}=E_k\boldsymbol{\psi}$ for ${\bf H}={\bf T}+{\bf V}\in \mathbf{M}(\mathcal{R})$, $\mathbf K(\boldsymbol{\psi})$ has $k-1$ negative eigenvalues and $M-k+2$ positive eigenvalues.

\textbf{Proof:}
Assuming that $\boldsymbol{\psi}$ is an eigenstate of ${\bf H}={\bf T}+{\bf V}$ with eigenvalue $\lambda$, then $\mathbf{H}^{(\lambda)}=\mathbf{T}-(\lambda \mathbf1 -\mathbf V)= \mathbf T- \mathbf D$ has $\boldsymbol{\psi}$ in its null space. Rearranging, $(\mathbf T+\mathbf D)\boldsymbol{\psi}=0$ implies $\sum_k T_{jk}\psi_k=D_j\psi_j$. Before using the definition of the force-balance equation, it is important to note that eigenvectors of symmetric matrices are real. Hence, for a single-particle in the eigenstate $\boldsymbol{\psi}$:
\begin{eqnarray}
K_{ij}&=&-2T_{ij}\psi_i\psi_j +2\delta_{ij} \psi_j \left(\sum_k T_{jk}\psi_k\right)\\
&=&-2T_{ij}\psi_i\psi_j-2\delta_{ij}D_j\psi_j^2\\
&=&\sum_{mn}(\delta_{mi}\psi_i)2(T_{mn}-\delta_{mn}D_n)(\delta_{nj}\psi_j)\\
\end{eqnarray}
Since $\psi$ has full support, $S_{mj}=\delta_{mj}\psi_j$ is non-singular and the number of $(+/0/-)$ eigenvalues are the same for $\mathbf{H}^{(\lambda)}$ and $\mathbf K(\boldsymbol{\psi})$ by Sylvester's theorem~\cite{Horn05}.
$\square$

According to numerical tests, if $\bos{\psi}$ does not have full support then $\mathbf K$ will have an additional vector in the null space.  This is consistent with the theorems from Ref.~\cite{Farzanehpour12}. The interacting extension of the present theorem does not seem to hold although we found that the many-body non-interacting ground state gives rise to $\mathbf K$ with the same inertia as $\mathbf{H}^{(\lambda_0)}$ consistent with the previous findings \cite{Farzanehpour12}.

\section{Conclusions}

In this article, we have described our software to implement potential inversion for a given time trace of quantum simulation data. Given the time-dependent density and an appropriate initial state, our method solves for the scalar Kohn-Sham potential.  In this article, we have discussed the implementation details, a heuristic for preparing initial multi-electrons states with non-zero momentum, and preliminary results for one- and two-electron examples.



Next steps for extending our implementation are performing comparisons and combinations with the implicit fix-point methods~\cite{Ruggenthaler11,Ruggenthaler12,Nielsen13} and methods using Kohn-Sham vector potentials \cite{Hodgson2013,Hodgson2016}.  We would also like to test our initial state preparation methods for more multi-electron systems.  

An interesting open question concerns the possible formation of shockwaves within our two-electron example.  Our method fails to continue the solution past the formation of a sharp feature which we suspect may be a cusp.  Further investigations are warranted using tools from the partial differential equations literature. In our present work, we are unable to observe non-analytic spatial behavior due to the lattice formulation.  Thus, an approach directly in in the continuum limit formulations could help address analytic behavior of this particular model.


We are also interested in connecting this work to the on-going excitement in the area of quantum computing.  Previously, we have shown that the $V$-representation problem can be solved efficiently using a quantum computer but it still remains an open question if all quantum computations remains efficiently simulatable with TDDFT given access to efficient solutions to the $V$-representation problem. Our next steps will also include using simulation data extracted from cloud-accessible quantum computers.

\section*{Acknowledgements}
This work was primarily supported by the U.S. Department of Energy, Office
of Science, Office of Advanced Scientific Computing Research, under the Quantum
Computing Application Teams program (Award 1979657).  JDW was also supported 
by Walter and Constance Burke award, the NSF (Grant 1820747) and 
additional funding from the DOE (Award A053685).

\bibliography{citations}
\end{document}